\documentclass[prl,amsmath,amsfonts,amssymb,letterpaper,superscriptaddress,twocolumn,floatfix%
]{revtex4-1}
\usepackage{bm}
\usepackage{graphicx}
\usepackage{color}

\def\citenum#1{{\def\@cite##1##2{##1}\cite{#1}}}

\newcommand\arrow\vec

\def\vec#1{\bm{#1}}

\def\negspace{\!}

\def\lrsub#1#2#3{{\vphantom{#1}}_{#2} \negspace {#1} \negspace {\vphantom{#1}}_{#3}}

\def\bra#1{\left\langle {#1} \right\rvert}
\def\ket#1{\left\lvert {#1} \right\rangle}

\def\inprod#1#2{\left\langle {#1} | {#2} \right\rangle}

\def\inprodsubsub#1#2#3#4{\lrsub {\inprod{#1}{#2}} {#3} {#4}}

\def\pqinprod#1#2{\inprodsubsub{#1}{#2} p q}
\def\qpinprod#1#2{\inprodsubsub{#1}{#2} q p}
\def\pinprod#1#2{\inprodsubsub{#1}{#2} p p}
\def\qinprod#1#2{\inprodsubsub{#1}{#2} q q}
\def\pqbraket\pqinprod
\def\qpbraket\qpinprod
\def\pbraket\pinprod
\def\qbraket\qinprod

\def\outprod#1#2{\ket {#1}\!\bra {#2}}

\def\1{I}

\def\v0{{\bvec 0}}

\begin{document}

\title{Entangling the optical frequency comb: simultaneous generation of multiple $\mathbf{2\times 2}$ and $\mathbf{2\times 3}$ continuous-variable cluster states in a single optical parametric oscillator}

\author{Hussain Zaidi}
\affiliation{Department of Physics, University of Virginia, 382 McCormick Road, Charlottesville, Virginia 22904-4714, USA}

\author{Nicolas C. Menicucci}
\email{nmen@princeton.edu}
\affiliation{Department of Physics, Princeton University, Princeton, NJ 08544, USA}
\affiliation{Department of Physics, The University of Queensland, Brisbane, Queensland 4072, Australia}

\author{Steven T. Flammia}
\affiliation{Perimeter Institute for Theoretical Physics, Waterloo, Ontario, N2L 2Y5 Canada}

\author{Russell Bloomer}
\affiliation{Department of Physics, University of Virginia, 382 McCormick Road, Charlottesville, Virginia 22904-4714, USA}

\author{Matthew Pysher}
\affiliation{Department of Physics, University of Virginia, 382 McCormick Road, Charlottesville, Virginia 22904-4714, USA}

\author{Olivier Pfister}
\email{opfister@virginia.edu}
\affiliation{Department of Physics, University of Virginia, 382 McCormick Road, Charlottesville, Virginia 22904-4714, USA}

\date{October~25, 2007}

\begin{abstract}

We report on our research effort to generate large-scale multipartite optical-mode entanglement using as few physical resources as possible.  We have previously shown that cluster- and GHZ-type $N$-partite continuous-variable entanglement can be obtained in an optical resonator that contains a suitably designed second-order nonlinear optical medium, pumped by at most ${\cal O}(N^2)$ fields. In this paper, we show that the frequency comb of such a resonator can be entangled into an arbitrary number of independent $2\times 2$ and $2\times 3$ continuous-variable cluster states by a single optical parametric oscillator pumped by just a few optical modes.
\end{abstract}

\maketitle


\section{Introduction}

Continuous-variable (CV) quantum information~\cite{Braunstein2005a} presents interesting distinctive features, first and foremost its well developed experimental interface by use of quantum optics, where the field amplitude and phase quadratures are the exact mathematical analogs of position and momentum of a quantum harmonic oscillator. In particular, the Bloch-Messiah reduction~\cite{Braunstein2005} provides an elegant method for generating any $N$-partite entangled CV quantum state having a Gaussian Wigner function by the interference of $N$ field modes emitted by $N$ independent single-mode squeezers. Examples have been given for CV Greenberger-Horne-Zeilinger (GHZ) states~\cite{vanLoock2000} and, more recently, for CV cluster states~\cite{Zhang2006,vanLoock2006}.

Cluster states, originally described using qubits~\cite{Briegel2001,Raussendorf2001} but later extended to qudits~\cite{Zhou2003} and continuous variables~\cite{Menicucci2006}, are the core resource for one-way quantum computing, a paradigm of quantum computation in which the unitary evolution of a quantum algorithm is implemented via measurements on a highly entangled state of a particular type (a cluster state), in a similar way to the use of quantum teleportation as a quantum computing primitive~\cite{Gottesman1999,Jozsa2005}. In addition, cluster states of any dimension possess remarkable entanglement resilience to measurement: when one measures a subsystem of a cluster state in an appropriate basis, the rest of the system stays entangled in a cluster state.  Cluster states are locally equivalent to Bell states for two entangled systems and to GHZ states for three entangled systems but become locally nonequivalent to those for four or more entangled systems. Recently, Grover's algorithm was implemented on four qubits using an optical one-way quantum computer~\cite{Walther2005}.

A one-way CV quantum computer based on Gaussian cluster states can perform universal quantum computing provided that at least one non-Gaussian detection operation is used~\cite{Lloyd1999,Menicucci2006}. Such a detection operation could be a photon-number resolving stage, towards which significant progress has recently been made~\cite{Gansen2007}, with the other stages being straightforward high-fidelity homodyne detection. Therefore, striving toward the large-scale quantum engineering of Gaussian states seems a particularly worthwhile endeavor.

In recent work~\cite{Pfister2004,Menicucci2007}, we have shown that the aforementioned $N$ independent squeezers and ${\cal O}(N^2)$ beamsplitters can be replaced with a {\em single} $N$-mode squeezer, based on a nonlinear photonic crystal (a quasi-phase-matched~\cite{Armstrong1962,Fejer1992,Lifshitz2005} periodically poled~\cite{Myers1995} ferroelectric) pumped by ${\cal O}(N^2)$ field modes, and no output interferometer. We outlined detailed experimental proposals for the generation of CV tri- and quadripartite GHZ~\cite{Pfister2004,Bradley2005} and cluster~\cite{Menicucci2007} states. In these works, the entangled systems are different eigenmodes of the optical parametric oscillator (OPO) cavity, labeled by their resonant frequencies and polarizations. Such a physical system exhibits high classical coherence, as spectacularly illustrated by the use of optical frequency combs consisting of $\sim 10^6$ locked oscillating modes of a femtosecond laser as a scale-universal time/frequency standard~\cite{Hall2006,Hansch2006}. Recently, ultrastable phase-locked OPOs were used to observe macroscopic Hong-Ou-Mandel interference~\cite{Feng2004} and bipartite CV entanglement~\cite{Jing2006}. The question we ask here is, can the optical frequency comb be turned into a quantum computer register?

The theoretical answer is yes~\cite{Pfister2004,Menicucci2007}. However, while there is no fundamental limit to the generalization of CV multipartite entanglement to the whole frequency comb, the experimental implementation does face one hurdle: the consequences of unwanted nonlinear interactions of modes within the cluster with others outside the cluster. These consequences are difficult to assess precisely because of the absence of good multipartite entanglement measures, but can be easily seen to be detrimental. We therefore adopt the conservative approach of designing the $N$-mode squeezer perfectly---i.e., with only the desired couplings between the modes inside the entangled set. The use of the selection rules provided by additional quantum numbers, such as polarization, is a logical avenue to solve this problem. However, the two-dimensional polarization basis does not offer enough degrees of freedom to scale experimental implementations beyond 4-mode entanglement. Nevertheless, we show in this paper that the simultaneous generation of multiple, larger CV cluster states is possible in the frequency comb of a single OPO, using only a few pump fields and simple nonlinear media.

\section{Multipartite entanglement in the optical frequency comb: experimental outlook}

In quantum optics, the set of resonant modes of an optical cavity, analog to the harmonics of a vibrating string, can be seen as a set of quantum harmonic oscillators whose frequencies are equally spaced, in the absence of dispersion, by the free spectral range (a ``harmonic quantum harp"). If a nonlinear medium is inserted in the cavity and pumped by a monochromatic mode at frequency $\omega_{\text{pump}}$, pairs of modes $(m,n)$  of respective frequencies that $\omega_m+\omega_n=\omega_{\text{pump}}$ (temporal phase-matching condition, or photon energy conservation) will become coupled (Fig.~\ref{2pairs}). This coupling manifests itself by the creation or annihilation of a photon pair with one photon in mode $m$ and one photon in mode $n$. It is well-known that such an OPO yields bipartite entanglement below~\cite{Ou1992a,Schori2002} and above~\cite{Villar2005,Su2006,Jing2006} the OPO threshold. The principle of the method lies in considering whether an appropriate network of such bipartite couplings may lead to multipartite entanglement. As already mentioned, solutions always exist for GHZ and cluster states, although they are far from being unique~\cite{Pfister2004,Menicucci2007}. In the following, we give a detailed description of the method. 

\subsection{$\mathbf H$ (Hamiltonian)-graph states: physical description}

\begin{figure}[!htb]
\begin{center}
\includegraphics[width= 0.4 \columnwidth]{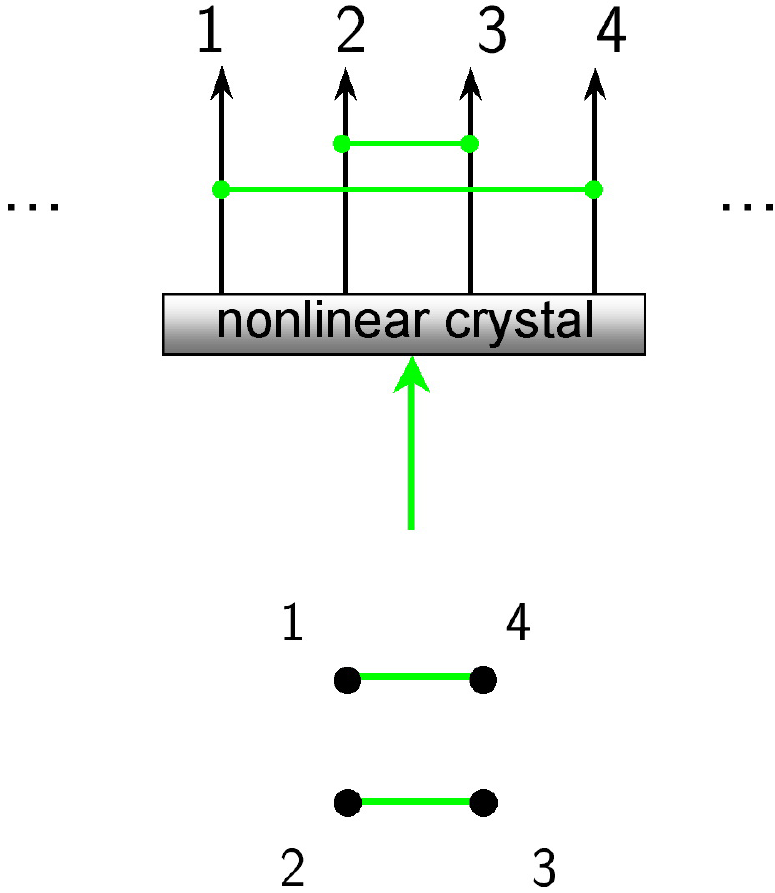}
\caption{\em  Physical system and corresponding $\cal H$-graph for a single pump mode.  The flow of time is from bottom to top of the figure. The horizontal direction is the optical frequency axis for the signal OPO modes, with the pump modes denoted by half their frequency. The $\cal H$-graph is drawn below and corresponds to mere pairwise entanglement.}
\label{2pairs}
\end{center}
\vspace{-.25in}
\end{figure}

The aforementioned Hamiltonian is of the form
\begin{align}
\label{eq:HTMS}
	\mathcal H = {i\hbar\kappa} \sum_{m,n} G_{mn}(a_m^\dag a_n^\dag - a_m a_n),
\end{align}
where $\kappa$ is an overall coupling strength, $a_m$ is the photon annihilation operator for cavity mode $m$, and $G=(G_{mn})$ is a square symmetric matrix describing the Hamiltonian coupling network: we make the hypothesis of equal coupling constants for all coupled modes, so that the elements of $G$ are either 0 or 1.  (We will relax this later to allow $-1$ as well.)  The matrix $G$ is thus the {\em adjacency matrix} of a graph representing $\mathcal H$, which we refer to as the $\mathcal H$-graph as a reminder of this correspondence.  An $\cal H$-graph has vertices denoting the field modes (in a vacuum state) and edges representing the non-zero terms of $\mathcal H$. The $\cal H$-graph allows us to easily visualize the couplings associated with the implementation of the Hamiltonian in Eq.~(\ref{eq:HTMS}).  We sometimes refer to states created by interactions of the form of Eq.~(\ref{eq:HTMS}) as $\cal H$-graph states, or even simply $\cal H$-graphs if this will not lead to confusion with the actual graph itself. Fig.~\ref{2pairs} displays the example of a monochromatically pumped (type-I) OPO, which can only generate entangled pairs. The $\cal H$-graph adjacency matrix in this case, denoted $G_1$, has a constant main skew diagonal:
\begin{equation}
G_1=\left(\begin{array}{cccc}
 0 & 0  & 0 & 1 \\
0 & 0  & 1&0\\
 0 & 1 & 0& 0 \\
 1 & 0 & 0  & 0 
\end{array}\right).
\end{equation}

In our recent work~\cite{Menicucci2007}, $G$ was constrained to be symmetric, real, and full rank. In this paper, we give more emphasis to the experimental viewpoint and add another constraint. Since the optical modes are labeled by their frequencies, the temporal phase-matching relation ($\omega_m+\omega_n = \omega_{\text{pump}}$) together with the assumption of constant interaction strength yield an adjacency matrix $G$ with constant skew-diagonals, that is, with ones at $m+n=\text{(const.)}$ for each given pump mode. Any matrix with constant skew-diagonals is called a Hankel matrix.  For the sake of experimental simplicity, we therefore restrict the discussion in this paper to $G$'s of Hankel form, even though more general ones can be considered (and may also be feasible)~\cite{Menicucci2007}. 

Throughout the rest of the paper, we adopt a more compact notation for Hankel matrices, in which we only list the entries along the top and down the right side of the matrix, with the top-right entry set off with slashes:
\begin{equation}
G_1 =[0,0,0/1/0,0,0].
\end{equation}
This shorthand provides a way to read off the experimental implementation immediately in the case of a cavity with an evenly spaced frequency comb of modes (and no polarization dependence): each nonzero entry in this ``vector'' corresponds to a pump beam.  The numerical value of the entry is proportional to the amplitude of the pump (negative would indicate a $\pi$-phase shift), and pumps corresponding to adjacent entries are apart in frequency by twice the free spectral range of the OPO cavity. 

\subsection{Square-cluster OPO}

\begin{figure}[!htb]
\begin{center}
\includegraphics[width= 0.4 \columnwidth]{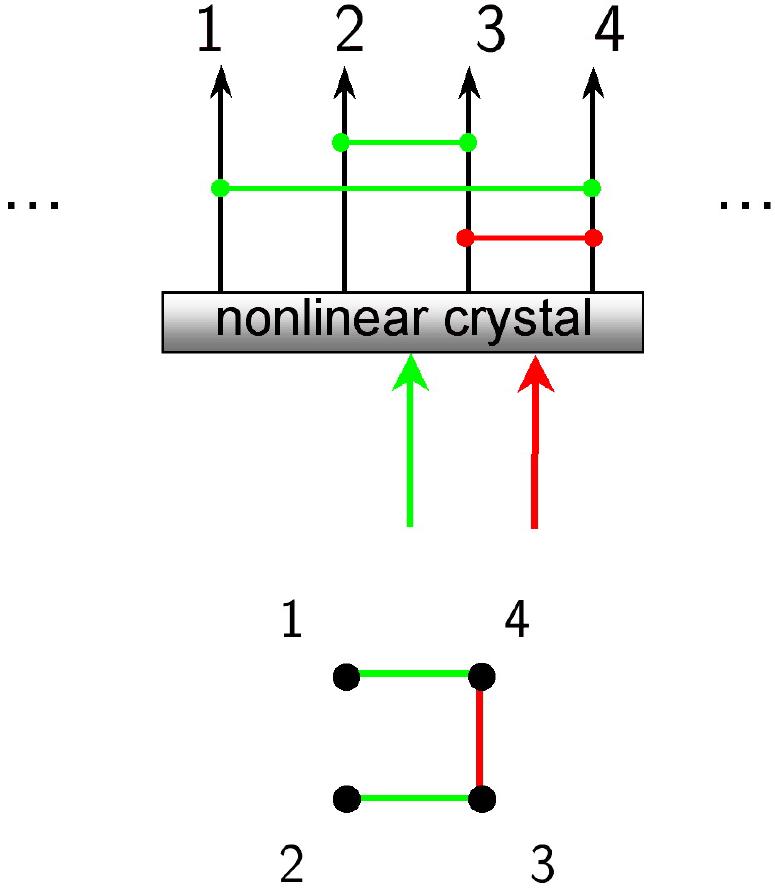}
\caption{\em  Physical system and corresponding $\cal H$-graph for a bimodal pump. Here, the interactions connect four modes each. As is shown in the text, these form a square cluster state. }
\label{1square}
\end{center}
\vspace{-.25in}
\end{figure}

Unlike qubit graph states (which include square-lattice cluster states)~\cite{Hein2006}, the entanglement properties of $\cal H$-graphs have not yet been studied in detail, except for our recent existence proof of an analytic correspondence between $\cal H$-graph and cluster states~\cite{Menicucci2007}. It is, however, remarkable that very simple $\cal H$-graphs on the frequency comb may yield fairly elaborate cluster states. Consider the physical system of Fig.~\ref{1square}, where two pump modes interact in the nonlinear crystal to yield four interacting modes. 
Note that it is crucial that the red connections be limited to only the one drawn in the figure. For example, if there were a fifth mode in the cavity, modes~2 and~5 (not drawn) must not be coupled by the pump mode drawn in red. This can be achieved by engineering the nonlinear optical interactions in quasi-phase-matched materials~\cite{Fejer1992,Lifshitz2005}, such as periodically poled $\rm KTiOPO_4$ (PPKTP)~\cite{Pooser2005}. Referring to Fig.\ref{1square} (and remembering that pumps are drawn at half their true frequency), we can write down the compact form of the adjacency matrix by inspection: 
\begin{equation}\label{g2}
G_2=[0,0,0/1/0,1,0].
\end{equation}
Under these conditions, it can then be shown that the 4-mode set actually generates a CV cluster state with a square graph. More precisely, the defining relation~\cite{vanLoock2006,Menicucci2007} for CV cluster states is satisfied, 
\begin{equation}
	\vec p - A\vec q\rightarrow \vec 0,
\end{equation}
where $\vec p$ and $\vec q$ are respective column vectors of quadratures $P_j=i(a^\dagger_j-a_j)$ and $Q_j=a_j+a^\dagger_j$ for each field mode $j$, $A$ is the adjacency matrix for a (possibly weighted) graph, and the arrow denotes the limit of large squeezing. Note that the source of entanglement is a {\em single} OPO~\cite{Menicucci2007}. Moreover, the present experimental proposal is simpler than that of Ref.~\citenum{Menicucci2007}.

A general proof of the above correspondence is based on the aforementioned analytical link which we will see in the next section.  We now give a particular-case proof of our claim that the $\cal H$-graph created by $G_2$ corresponds to a CV cluster state. Solving the Heisenberg equations for mode evolution under Eqs.~\eqref{eq:HTMS} and~\eqref{g2}, we obtain the following squeezed joint operators:
\begin{eqnarray}
\label{MMS1square}
\left(-\delta_-Q_1+ \delta_-Q_2 -Q_3 +Q_4\right)e^{-r\delta_+}\;, \\
\left(-\delta_+Q_1- \delta_+Q_2+Q_3+Q_4\right)e^{-r\delta_ -}\;, \\
\left(\delta_-P_1+\delta_-P_2+P_3+P_4\right)e^{-r\delta_ +}\;, \\
\left(\delta_+P_1-\delta_+P_2 -P_3+P_4\right)e^{-r\delta_ -}\;,
\label{MMS1square2}
\end{eqnarray}
where $r=\kappa t$ and $\delta_{\pm}=(\sqrt 5 \pm 1)/2$. Next, we compare these to the cluster-state equation, $\vec p - A\vec q\rightarrow \vec 0$, using the adjacency matrix $A$ given by
\begin{equation}
A = \left(
\begin{array}{cccc}
0 & 0 & 1  & \sqrt 5  \\
0 & 0 & \sqrt 5  & 1  \\
1& \sqrt 5 & 0  & 0  \\
\sqrt 5 & 1 & 0  & 0  
\end{array}
\right)\;,
\end{equation}
which corresponds to a weighted square graph with a certain choice of weightings. This yields
\begin{eqnarray}
\label{entwit1square}
\left(P_1 -\frac{Q_3}2 - \frac{\sqrt 5}2 Q_4\right) \rightarrow 0\\
\left(P_2 -\frac{\sqrt 5}2 Q_3 - \frac{Q_4}2\right) \rightarrow 0\\
\left(P_3 -\frac{Q_1}2 -\frac{\sqrt 5}2 Q_2\right) \rightarrow 0 \\
\left(P_4 -\frac{\sqrt 5}2 Q_1 - \frac{Q_2}2\right) \rightarrow 0
\label{entwit1square2}
\end{eqnarray}
which can be easily seen to be equivalent to Eqs.~(\ref{MMS1square}--\ref{MMS1square2}), in the limit $r\rightarrow 0$,  if we rotate modes 3 and 4 by $\pi/2$ ($Q\mapsto P$, $P\mapsto -Q$). This completes the proof.

\subsection{Multiple-square-cluster OPO}

\subsubsection{Principle and first experimental implementation}

\begin{figure}[!htb]
\begin{center}
\includegraphics[width= 0.9 \columnwidth]{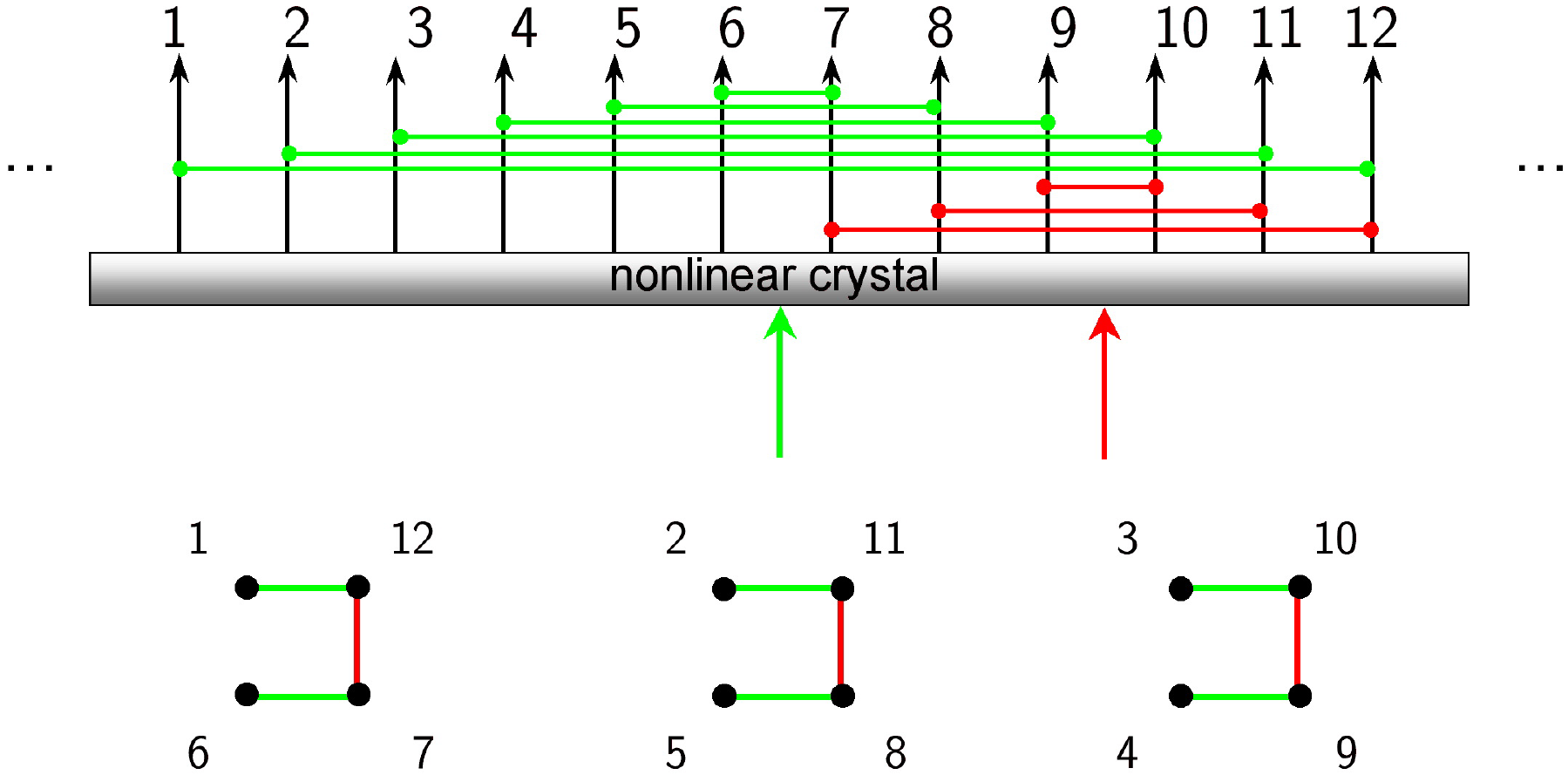}
\caption{\em Physical system and corresponding $\cal H$-graph for a bimodal pump. Here, the interactions define three (or more) sets of four modes each. As is shown in the text, each of these form a square cluster state.}
\label{manysquares}
\end{center}
\vspace{-.25in}
\end{figure}

As seen in Fig.~\ref{2pairs}, a single pump mode can generate an arbitrary number of entangled pairs in the same OPO, at least within the phase-matching bandwidth, which can be made orders of magnitude larger than the OPO's free spectral range. This somewhat trivial fact can be generalized to the square cluster of Fig.\ref{1square}: with the same number of pump modes (two), an {\it arbitrary number} of square clusters can actually be generated, as depicted in Fig.~\ref{manysquares} for three copies. 
This corresponds to a Hankel matrix of the same pattern as $G_2$ but for $4N$ modes rather than 4.  Once again, we can immediately read off the Hankel-shorthand form of $G_3$ by inspection of the diagram:
\begin{align}
\label{g3}
G_3= [0_{11}/1/ 0_5,1,0_5]\;,
\end{align}
where $0_n$ is shorthand for a string of $n$ zeros.  From an experimental perspective, the few pump modes and single type of nonlinear interaction greatly simplify the setup.  One still has to prevent the red pump from introducing spurious mode couplings (i.e., couplings of the indicated modes to other modes outside the set), for example by designing sharp quasiphasematching cutoffs in the nonlinear crystal or by  filtering mode amplitudes or engineering dispersion in the OPO cavity.  Such a constraint isn't needed for the green pump because any spurious couplings generated would be strictly outside of the desired mode set.

\subsubsection{Second experimental implementation}

\begin{figure}[!htb]
\begin{center}
\includegraphics[width= 0.9 \columnwidth]{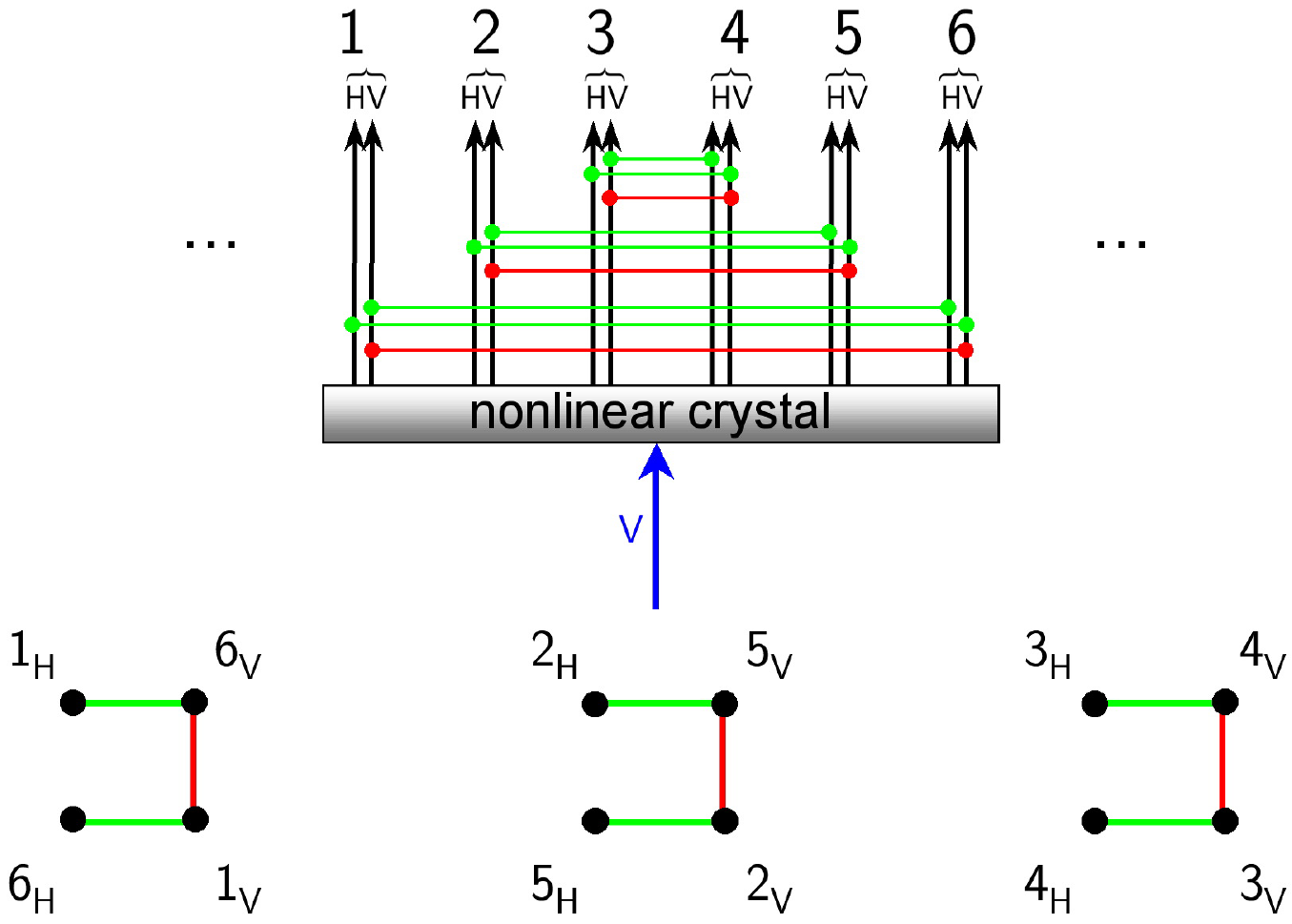}
\caption{\em Physical system and corresponding $\cal H$-graph for a single pump mode. The letters $H$ and $V$ denote the optical polarizations of each mode, which are frequency degenerate. Here, the interactions define three (or more) sets of four modes, each of which form a square cluster state.}
\label{manysquares2}
\end{center}
\vspace{-.25in}
\end{figure}

Note that this experimental generation of multiple square clusters is far from unique and can be even more efficiently and conveniently implemented by use of the polarization degree of freedom, as depicted in Fig.~\ref{manysquares2}. 
This yields several advantages: first, one uses but a single pump mode; second, one has the advantage of connecting entangled modes in {\em closed} sets, without any additional engineering of the nonlinear interaction bandwidths. The interactions yield exactly the same equations as Eqs.~(\ref{entwit1square}--\ref{entwit1square2}).

The experimental implementation is relatively straightforward: one can use one $ZZZ$ crystal placed in sequence with a $YZY$ crystal (first letter denotes pump polarization and other two the signal polarizations), with one crystal rotated at 90 degrees from the other. In this case, a vertically polarized pump generates the necessary $VHV$, $VVH$, and $VVV$ interactions. Moreover, if the crystals have the same length, the fact they are rotated by 90 degrees from each other ensures that the frequency combs corresponding to each polarization will have the same free spectral range, which in turns ensures common resonance conditions at a given cavity length. 

\subsubsection{Third experimental implementation}

\begin{figure}[!htb]
\begin{center}
\includegraphics[width= 0.9 \columnwidth]{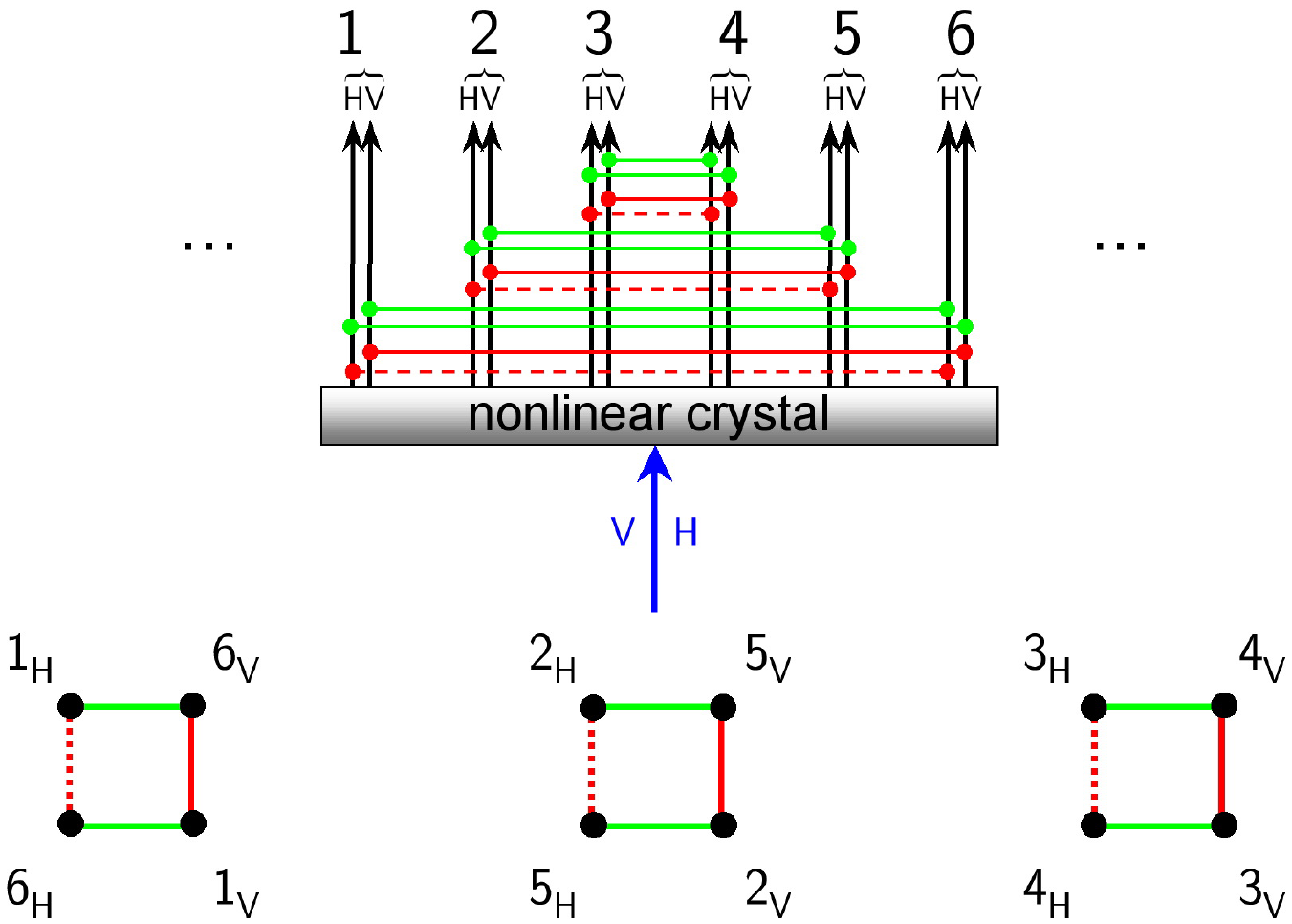}
\caption{\em Physical system and corresponding $\cal H$-graph for two frequency degenerate pump modes. Here, the interactions define three (or more) sets of four modes, each of which form a square cluster state~\cite{Menicucci2007}. Note that the opposite interaction (dashed line) is not necessary but gives a more balanced cluster state.}
\label{manysquares3}
\end{center}
\vspace{-.25in}
\end{figure}

A variant of the previous implementation allows for generation of multiple copies of the exact proposal of Ref.~\citenum{Menicucci2007}, which yields a more balanced square cluster state. This is achieved by adding an additional interaction (Fig.~\ref{manysquares3}, dashed lines) of opposite effect to the three former ones (i.e.\ upconverting if the three are downconverting, or vice versa). This can be ensured by seeding the OPO below threshold with coherent-state signals, appropriately phase-shifted with respect to the pump fields. 
The experimental implementation consists, for example, in using 2 PPKTP crystals: one that simultaneously quasiphasematches the $YZY$ and $ZZZ$ interactions~\cite{Pooser2005} and one only the $ZZZ$ interaction, rotated at 90 degrees from the other. In this case, vertically and horizontally polarized pumps generate interactions $VHV$ and $VVH$, $VVV$, and the opposite $HHH$. 

The resulting lower weighting~\cite{Menicucci2007} can be seen, as before, by solving the quantum evolution for the multimode squeezing, to be compared to Eqs.(\ref{MMS1square}--\ref{MMS1square2}),
\begin{eqnarray}
\label{MMSmanysquares}
\left(Q_1 + Q_2 -\sqrt 2\ Q_4\right)\ e^{-r\sqrt 2}\\ 
\left(Q_1 - Q_2 + \sqrt 2\ Q_3\right)\ e^{-r\sqrt 2}\\ 
\left(P_1 + P_2 + \sqrt 2\ P_4 \right)\ e^{-r\sqrt 2}\\ 
\left(P_1 - P_2 -\sqrt 2\ P_3\right)\ e^{-r\sqrt 2}, 
\label{MMSmanysquares2}
\end{eqnarray}
 and deducing, as before, the following cluster state relations, to be compared with Eqs.(\ref{entwit1square}--\ref{entwit1square2}):
 \begin{eqnarray}
\label{entwitmanysquares}
\left(\sqrt 2 P_1 + Q_3 - Q_4\right) \rightarrow\ 0\\ 
\left(\sqrt 2 P_2 - Q_3 + Q_4\right) \rightarrow\ 0\\ 
\left(\sqrt 2 P_3 + Q_1 - Q_2\right) \rightarrow\ 0\\ 
\left(\sqrt 2 P_4 - Q_1 + Q_2\right) \rightarrow\ 0.
\label{entwitmanysquares2}
\end{eqnarray}
These implementations are currently underway at the University of Virginia.

These methods therefore yield an {\em arbitrary number} of disconnected 4-mode square cluster states {\em from just a single OPO}, limited only by the number of modes within the phase-matching bandwidth. The whole frequency comb can thus be partitioned into entangled subsets by using only one or two pump modes. Our eventual goal, of course, is an arbitrarily large square-lattice grid (or other universal graph), and many disconnected squares isn't the same at all.  Still, the fact that only one or two pumps are needed to generate so many copies of this (rather modest) cluster state leaves us hopeful that this is a worthwhile line of research to pursue.  In the next section, we give the theoretical basis for this result and show that this process can also be used for 8-mode cubic cluster states.  These can each be converted, if desired, into $2\times 3$ square-lattice cluster states at the expense of homodyne detection on two modes per cube.


\section{Multipartite entanglement in the optical frequency comb: theoretical outlook}

\subsection{Mathematical relation between H- and cluster graphs}

The relation between a bipartite CV cluster state, represented by the adjacency matrix
\begin{align}
\label{eq:AfromA0}
	A =
	\begin{pmatrix}
		0 		& A_0  \\
		A_0^T	& 0
	\end{pmatrix}
	\;,
\end{align}
and the aforementioned $\cal H$-graph matrix $G$ is given by~\cite{Menicucci2007}
\begin{align}
\label{eq:Ggeneral}
	G &=
	\begin{pmatrix}
		[B - A_0 C A_0^T]		& [B A_0 + A_0 C]	\\
		[C A_0^T + A_0^T B]		& [A_0^T B A_0 - C]
	\end{pmatrix}
	\;.
\end{align}
where $B$ and $C$ are arbitrary symmetric positive-definite matrices. This relation allows for abundant (in fact, excessive) freedom in the choice of $G$ for a given $A$.  For concreteness, let's take $B = C = \tfrac I 2$, where $I$ is the identity matrix.  Now assume that $A$ is unitary, i.e. $A_0A_0^T = A_0^TA_0 = I$.  Then Eq.~\eqref{eq:Ggeneral} yields
\begin{align}
\label{eq:Gspecific}
	G =
	\begin{pmatrix}
		0		& A_0	\\
		A_0^T		& 0
	\end{pmatrix}
	= A
	\;.
\end{align}
Restricting to unitary $A$'s therefore can be used to eliminate the distinction between CV graphs and $\cal H$-graphs, simplifying the problem greatly (at the expense of some generality).  Keeping in mind the possible limitations of such a restriction, we focus only on unitary $A$'s.  Using $\cong$ to indicate equality up to a renumbering of modes, we then inquire whether we can always find a $G \cong A$, where $G$ is in Hankel form, for obvious experimental convenience.

\subsection{Simultaneously generating multiple copies of a CV cluster state}

For reasons that will become clear shortly, let's begin with a bipartite graph having a unitary, Hankel $A_0$.  Since $A_0$ is Hankel, it is automatically symmetric.  Thus, Eq.~\eqref{eq:AfromA0} can be expressed as $A = F_2 \otimes A_0$, where
\begin{align}
\label{eq:Fn}
	F_n = [0_{n-1}/1/0_{n-1}]
\end{align}
is the $n \times n$ skew-identity.  Since $A_0$ is unitary, by the argument above, we can choose for our $\cal H$-graph,
\begin{align}
\label{eq:Gastensor}
	G = A_0 \otimes F_2 \cong F_2 \otimes A_0 = A\;,
\end{align}
since exchanging the order of a tensor product is equivalent to a renumbering of the modes.  It's straightforward to show that since $A_0$ is Hankel, then $A_0 \otimes F_n$ (for any~$n$) is also Hankel.  Considering that $F_{2N} \otimes A_0$ is zero except for $2N$ blocks of $A_0$ on the main skew-diagonal, it can be seen that it is the adjacency matrix for $N$ distinct copies of the graph corresponding to $A$.  Also, $F_{2N} \otimes A_0$ is unitary if $A_0$ is unitary.  Since $F_{2N} \otimes A_0 \cong A_0 \otimes F_{2N}$, the latter being Hankel if $A_0$ is Hankel, then for any unitary, Hankel $A_0$, we know that
\begin{align}
\label{eq:GforNcopies}
	G_N = A_0 \otimes F_{2N}
\end{align}
is Hankel and will create $N$ copies of the CV graph $A$ as defined in Eq.~\eqref{eq:AfromA0}.

\subsection{Simultaneous generation of $\mathbf{2\times 2}$ and $\mathbf{2 \times 3}$ cluster states}

To apply these results, let's first consider the case of multiple square clusters in a single OPO. A unitary $A_0$ for a single square cluster is~\cite{Menicucci2007}
\begin{align}
\label{eq:square}
	A_0 = \frac {1} {\sqrt{2}} \begin{pmatrix}
		-1	& 1	\\
		1	& 1
	\end{pmatrix}
	= \frac {1} {\sqrt{2}} [-1/1/1]\;,
\end{align}
which is also a Hankel matrix.  Using the results above, we have that
\begin{align}
\label{eq:Hankelmultisquare}
	G_N = A_0 \otimes F_{2N} = \frac {1} {\sqrt{2}} [\bar 0,-1,\bar 0/1/\bar 0,1,\bar 0]\;,
\end{align}
where $\bar 0 = 0_{2N-1}$, is a Hankel matrix for the $\cal H$-graph that will generate $N$ disconnected squares from a single OPO using just three pumps.

Note that the negative matrix elements in $G_N$ are required for unitarity but are absent in Eq.~\eqref{g3} in the experimental section, which yields a CV cluster state with the same graph but different weights.  As we previously indicated, requiring that $A$ be unitary is not necessary; we do it here only for the convenience of the calculations.  It proves useful, though, because we can extend this method to generate an arbitrary number of disconnected 8-mode cubic cluster states, each of which can be reduced by measurement to a $2 \times 3$ square-lattice cluster state, if desired.  We show this presently.

\begin{figure}[!htb]
\begin{center}
\includegraphics[width= 0.9 \columnwidth]{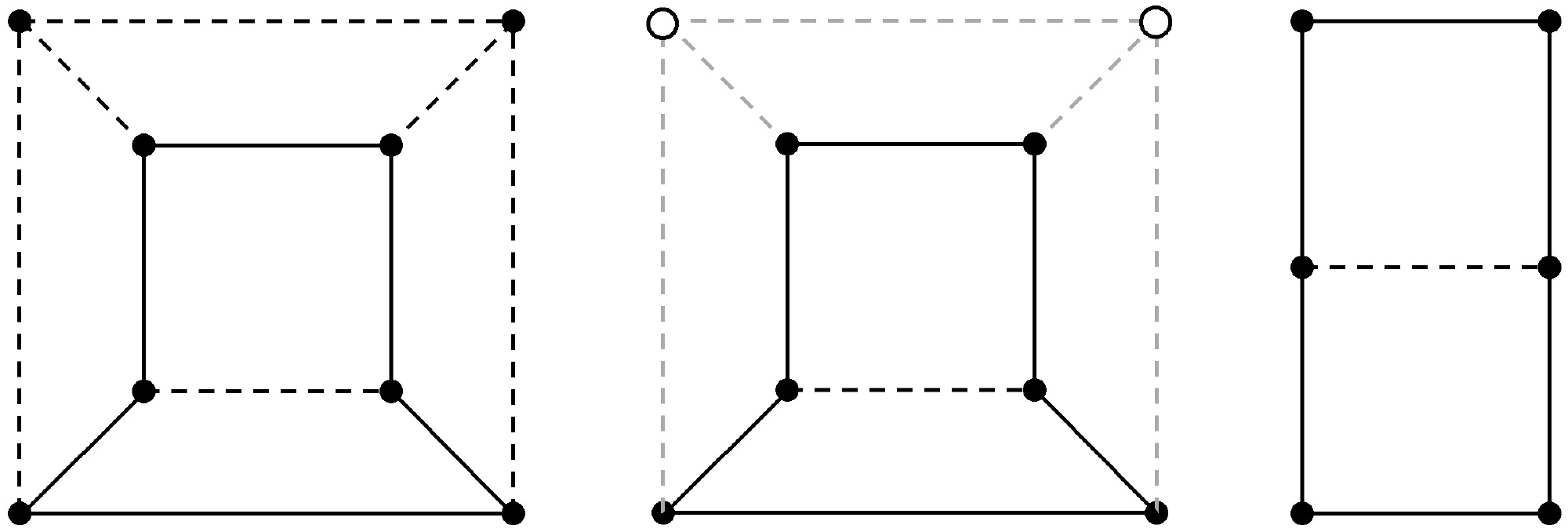}
\caption{\em Cubic cluster graph. Dashed lines indicate negative-weight edges.  Measurement of any two neighboring vertices (open circles) in the position (amplitude quadrature) basis disentangles those vertices from the rest of the graph, thereby yielding a $2\times 3$ cluster.}
\label{1cube}
\end{center}
\vspace{-.25in}
\end{figure}

The unitary form of $A_0$ for a cubic cluster is
\begin{align}
\label{eq:Hankelcube}
	A_0 = \frac {1} {\sqrt{3}} [-1,-1,1/0/1,1,-1]\;.
\end{align}
The corresponding graph is depicted in Fig.~\ref{1cube}. Note that a cubic cluster state reduces to a $2\times 3$ one under the measurement of two neighboring vertices in the position (amplitude quadrature) basis.
Using the results above, a generating $\cal H$-graph matrix in Hankel form for $N$ disconnected cubes is
\begin{align}
\label{eq:Hankelmultisquare2}
	G_N &= A_0 \otimes F_{2N} \nonumber \\
	&= \frac {1} {\sqrt{3}} [\bar 0,-1,\bar 0,-1,\bar 0,1,\bar 0/0/\bar 0,1,\bar 0,1,\bar 0,-1,\bar 0]\;,
\end{align}
where, again, $\bar 0 = 0_{2N-1}$.  We stress that this generation of an arbitrary number of disconnected cubes can be achieved using only six pumps and a single OPO.  The maximum number of cubes is limited only by the number of modes within the phase-matching bandwidth.


\section{Conclusion}

In this paper, we have expanded our previous results \cite{Menicucci2007} on the generation of CV cluster states with a minimum amount of physical resources, in particular by adding the constraint that the $\cal H$-graph adjacency matrix be of Hankel form and unitary. Although justifiably limiting the generality of the treatment, this constraint makes it easier to connect to concrete and simple experimental cases, for which we show that large sets of independent $2\times 2$ and $2\times 3$ square-grid cluster states can be obtained in a single OPO.  Multiple single-cube cluster states can also be produced in this fashion.

A remarkable feature of this sort of Hankel-unitary duplication procedure is that the number of pumps is independent of the number of squares or cubes being generated, as is the case for multiple entangled pairs; we need only three pumps at most (and as few as one, as shown in the experimental sections) to generate multiple squares and six pumps to generate multiple cubes. This is significantly better than our earlier estimate of the $O(N^2)$ pump fields needed to create a cluster with $N$ modes. This estimate was based on the conservative assumption of one pump per graph edge and $N(N-1)/2\sim N^2$ edges in a complete graph. We observe that square-lattice cluster states are significantly sparser, with $O(N)$ edges.  In addition, single pumps can couple many mode pairs when Hankel matrices are used.  Work is currently underway to experimentally generate these CV cluster states and to explore what other types of states can be created under these conditions.
  
We thank Luke Langsjoen for useful discussions and suggestions.  NCM acknowledges support from the U.S. Dept.\ of Defense and the National Science Foundation, STF from ONR Grant No.\ N00014-07-1-0304, and HZ, RB, MP, and OP from NSF Grants No.\ PHY-0555522 and No.\ CCF-0622100.


\end{document}